\documentstyle[twocolumn,aps,graphicx]{revtex}

\begin{document}
\twocolumn[\hsize\textwidth\columnwidth\hsize\csname
@twocolumnfalse\endcsname

\def\simge{\hspace*{0.2em}\raisebox{0.5ex}{$>$}
     \hspace{-0.8em}\raisebox{-0.3em}{$\sim$}\hspace*{0.2em}}
\def\simle{\hspace*{0.2em}\raisebox{0.5ex}{$<$}
     \hspace{-0.8em}\raisebox{-0.3em}{$\sim$}\hspace*{0.2em}}
\def\bra#1{{\langle#1\vert}}
\def\ket#1{{\vert#1\rangle}}
\def\coeff#1#2{{\scriptstyle{#1\over #2}}}
\def\undertext#1{{$\underline{\hbox{#1}}$}}
\def\hcal#1{{\hbox{\cal #1}}}
\def\sst#1{{\scriptscriptstyle #1}}
\def\eexp#1{{\hbox{e}^{#1}}}
\def\rbra#1{{\langle #1 \vert\!\vert}}
\def\rket#1{{\vert\!\vert #1\rangle}}
\def\lsim{{ <\atop\sim}}
\def\gsim{{ >\atop\sim}}
\def\nubar{{\bar\nu}}
\def\psibar{{\bar\psi}}
\def\Gmu{{G_\mu}}
\def\alr{{A_\sst{LR}}}
\def\wpv{{W^\sst{PV}}}
\def\evec{{\vec e}}
\def\notq{{\not\! q}}
\def\notk{{\not\! k}}
\def\notp{{\not\! p}}
\def\notpp{{\not\! p'}}
\def\notder{{\not\! \partial}}
\def\notcder{{\not\!\! D}}
\def\notA{{\not\!\! A}}
\def\notv{{\not\!\! v}}
\def\Jem{{J_\mu^{em}}}
\def\Jana{{J_{\mu 5}^{anapole}}}
\def\nue{{\nu_e}}
\def\mn{{m_\sst{N}}}
\def\mns{{m^2_\sst{N}}}
\def\me{{m_e}}
\def\mes{{m^2_e}}
\def\mq{{m_q}}
\def\mqs{{m_q^2}}
\def\mz{{M_\sst{Z}}}
\def\mzs{{M^2_\sst{Z}}}
\def\ubar{{\bar u}}
\def\dbar{{\bar d}}
\def\sbar{{\bar s}}
\def\qbar{{\bar q}}
\def\sstw{{\sin^2\theta_\sst{W}}}
\def\gv{{g_\sst{V}}}
\def\ga{{g_\sst{A}}}
\def\pv{{\vec p}}
\def\pvs{{{\vec p}^{\>2}}}
\def\ppv{{{\vec p}^{\>\prime}}}
\def\ppvs{{{\vec p}^{\>\prime\>2}}}
\def\qv{{\vec q}}
\def\qvs{{{\vec q}^{\>2}}}
\def\xv{{\vec x}}
\def\xpv{{{\vec x}^{\>\prime}}}
\def\yv{{\vec y}}
\def\tauv{{\vec\tau}}
\def\sigv{{\vec\sigma}}
\def\sst#1{{\scriptscriptstyle #1}}
\def\gpnn{{g_{\sst{NN}\pi}}}
\def\grnn{{g_{\sst{NN}\rho}}}
\def\gnnm{{g_\sst{NNM}}}
\def\hnnm{{h_\sst{NNM}}}

\def\xivz{{\xi_\sst{V}^{(0)}}}
\def\xivt{{\xi_\sst{V}^{(3)}}}
\def\xive{{\xi_\sst{V}^{(8)}}}
\def\xiaz{{\xi_\sst{A}^{(0)}}}
\def\xiat{{\xi_\sst{A}^{(3)}}}
\def\xiae{{\xi_\sst{A}^{(8)}}}
\def\xivtez{{\xi_\sst{V}^{T=0}}}
\def\xivteo{{\xi_\sst{V}^{T=1}}}
\def\xiatez{{\xi_\sst{A}^{T=0}}}
\def\xiateo{{\xi_\sst{A}^{T=1}}}
\def\xiva{{\xi_\sst{V,A}}}

\def\rvz{{R_\sst{V}^{(0)}}}
\def\rvt{{R_\sst{V}^{(3)}}}
\def\rve{{R_\sst{V}^{(8)}}}
\def\raz{{R_\sst{A}^{(0)}}}
\def\rat{{R_\sst{A}^{(3)}}}
\def\rae{{R_\sst{A}^{(8)}}}
\def\rvtez{{R_\sst{V}^{T=0}}}
\def\rvteo{{R_\sst{V}^{T=1}}}
\def\ratez{{R_\sst{A}^{T=0}}}
\def\rateo{{R_\sst{A}^{T=1}}}

\def\mro{{m_\rho}}
\def\mks{{m_\sst{K}^2}}
\def\mpi{{m_\pi}}
\def\mpis{{m_\pi^2}}
\def\mom{{m_\omega}}
\def\mphi{{m_\phi}}
\def\Qhat{{\hat Q}}

\def\FOS{{F_1^{(s)}}}
\def\FTS{{F_2^{(s)}}}
\def\GAS{{G_\sst{A}^{(s)}}}
\def\GES{{G_\sst{E}^{(s)}}}
\def\GMS{{G_\sst{M}^{(s)}}}
\def\GATEZ{{G_\sst{A}^{\sst{T}=0}}}
\def\GATEO{{G_\sst{A}^{\sst{T}=1}}}
\def\mdax{{M_\sst{A}}}
\def\mustr{{\mu_s}}
\def\rsstr{{r^2_s}}
\def\rhostr{{\rho_s}}
\def\GEG{{G_\sst{E}^\gamma}}
\def\GEZ{{G_\sst{E}^\sst{Z}}}
\def\GMG{{G_\sst{M}^\gamma}}
\def\GMZ{{G_\sst{M}^\sst{Z}}}
\def\GEn{{G_\sst{E}^n}}
\def\GEp{{G_\sst{E}^p}}
\def\GMn{{G_\sst{M}^n}}
\def\GMp{{G_\sst{M}^p}}
\def\GAp{{G_\sst{A}^p}}
\def\GAn{{G_\sst{A}^n}}
\def\GA{{G_\sst{A}}}
\def\GETEZ{{G_\sst{E}^{\sst{T}=0}}}
\def\GETEO{{G_\sst{E}^{\sst{T}=1}}}
\def\GMTEZ{{G_\sst{M}^{\sst{T}=0}}}
\def\GMTEO{{G_\sst{M}^{\sst{T}=1}}}
\def\lamd{{\lambda_\sst{D}^\sst{V}}}
\def\lamn{{\lambda_n}}
\def\lams{{\lambda_\sst{E}^{(s)}}}
\def\bvz{{\beta_\sst{V}^0}}
\def\bvo{{\beta_\sst{V}^1}}
\def\Gdip{{G_\sst{D}^\sst{V}}}
\def\GdipA{{G_\sst{D}^\sst{A}}}
\def\fks{{F_\sst{K}^{(s)}}}
\def\FIS{{F_i^{(s)}}}
\def\fpi{{F_\pi}}
\def\fk{{F_\sst{K}}}

\def\RAp{{R_\sst{A}^p}}
\def\RAn{{R_\sst{A}^n}}
\def\RVp{{R_\sst{V}^p}}
\def\RVn{{R_\sst{V}^n}}
\def\rva{{R_\sst{V,A}}}
\def\xbb{{x_B}}

\def\PR#1{{{\em   Phys. Rev.} {\bf #1} }}
\def\PRC#1{{{\em   Phys. Rev.} {\bf C#1} }}
\def\PRD#1{{{\em   Phys. Rev.} {\bf D#1} }}
\def\PRL#1{{{\em   Phys. Rev. Lett.} {\bf #1} }}
\def\NPA#1{{{\em   Nucl. Phys.} {\bf A#1} }}
\def\NPB#1{{{\em   Nucl. Phys.} {\bf B#1} }}
\def\AoP#1{{{\em   Ann. of Phys.} {\bf #1} }}
\def\PRp#1{{{\em   Phys. Reports} {\bf #1} }}
\def\PLB#1{{{\em   Phys. Lett.} {\bf B#1} }}
\def\ZPA#1{{{\em   Z. f\"ur Phys.} {\bf A#1} }}
\def\ZPC#1{{{\em   Z. f\"ur Phys.} {\bf C#1} }}
\def\etal{{{\em   et al.}}}

\def\delalr{{{delta\alr\over\alr}}}
\def\pbar{{\bar{p}}}
\def\lamchi{{\Lambda_\chi}}

\title{Charged Current Universality in the MSSM}

\author{A. Kurylov$^{a,b}$  and
M.J. Ramsey-Musolf$^{a,b}$
\\[0.3cm]
}
\address{
$^a$ Kellogg Radiation Laboratory, California Institute of Technology,
Pasadena, CA 91125\ USA\\
$^b$ Department of Physics, University of Connecticut, Storrs, CT 06269\ USA
}


\maketitle

\begin{abstract}

We compute the complete one-loop contributions to low-energy charged
current weak
interaction observables in the Minimal Supersymmetric Standard Model (MSSM). We
obtain the constraints on the MSSM parameter space which arise when precision
low-energy charged current data are analyzed in tandem with measurements of the
muon anomaly. While the data allow the presence of at least one light
neutralino,
they also imply a pattern of mass splittings among first and second generation
sleptons and squarks which contradict predictions of widely used models for
supersymmetry breaking mediation.

\end{abstract}



\vspace{0.3cm}
]

\pagenumbering{arabic}

The universality of the charged current  weak interaction (CCWI) is an
important feature of the
Standard Model (SM). The presence of a common coupling strength and
$(V-A)\times
(V-A)$ current-current interaction structure for all CCWI processes has
been tested with high precision in
a number of leptonic and semileptonic experiments. The results place
significant limits on scenarios for
physics beyond the SM which may generate breakdowns of CCWI universality.
To date, however, the
implications of universality tests for supersymmetric extensions of the SM
-- a leading candidate for
\lq\lq new physics" -- have not been fully explored.
Although supersymmetric theories which break R-parity conservation have
been considered\cite{barger,MRM00,dreiner}, no
analysis has been performed for the simplest version of supersymmetry
(SUSY): the Minimal Supersymmetric
Standard Model (MSSM). In this letter, we report on results obtained from
such an analysis.

Low energy SUSY is an attractive scenario from a number of standpoints: it
provides a solution to the
hierarchy problem associated with Higgs mass renormalization; it produces
coupling unification at the GUT
scale; and it is a prediction of superstring theory. It remains to be seen,
however, which version of SUSY
correctly describes electroweak phenomena, and as we argue below, low energy
precision measurements may provide important clues. In particular, details
of the superpartner spectrum
({\em e.g.}, masses and mixing angles) are largely unknown. Limits on
branching ratios obtained from
collider data provide, in general, only weak lower bounds. Low-energy CCWI
observables can provide
complementary information, since they effectively compare the relative
importance of effects produced by
different species or generations of superpartners.

In the MSSM, R-parity (and, thus, $B-L$) is
conserved, so that SUSY corrections to low energy observables arise only
via tiny loop effects
(non-conservation of R-parity allows for the presence of new tree-level
SUSY effects). In order to become
sensitive to such contributions, one generally requires a precision of
$\sim(\alpha/\pi)\times (M/{\tilde
M})^2$, where $M$ is the relevant mass of a SM particle and  ${\tilde M}$
is a superpartner mass. The
recent report\cite{Brown} of a 2.6$\sigma$ deviation of the muon anomaly
from the
SM prediction provides the first,
tantalizing hint of MSSM loop effects .
In this case, $M=m_\mu$, making the
ppm precision of the muon
anomaly measurement sensitive to superpartner masses on the order of a few
hundred GeV \cite{Everett}.

In contrast, for CCWI observables one has $M\sim M_W$, so that only a few
$\times 10^{-3}$ precision is needed to achieve comparable sensitivity. A
number of low-energy CCWI
measurements have achieved this level of precision, and others are poised
to do so in the near future.
These observables include: (a) the branching ratio
$R_{e/\mu}=\Gamma(\pi^+\to e^+\nu_e + \pi^+\to
e^+\nu_e\gamma)/\Gamma(\pi^+\to \mu^+\nu_\mu + \pi^+\to
\mu^+\nu_\mu\gamma)$; (b) the $ft$ values for
\lq\lq superallowed" $(J^\pi,I)=(0^+,0)\to (0^+,0)$ Fermi nuclear
$\beta$-decays; and (c) the neutron
lifetime, $\tau_n$. In addition, a more precise determination of $\tau_n$
is underway at NIST, as are
measurements of the parity-violating neutron $\beta$-decay parameter $A$ at
LANSCE and the $\pi^+$
$\beta$-decay branching ratio at PSI.


Of particular interest for our analysis are the
results of superallowed nuclear $\beta$-decays, from which one extracts the
Cabibbo-Kobayashi-Maskawa (CKM)
quark mixing matrix element $|V_{ud}|$. When $|V_{ud}|$ is considered along
with
the values of $|V_{us}|$ and
$|V_{ub}|$ determined from $K_{e3}$ and $B$-meson decays, respectively, one
obtains for the sum of the
squares a result falling below the unitarity requirement by
$2.2\sigma$\cite{hardy}. In what follows, we discuss the
implications of this deviation -- which will be tested via the new $\tau_n$
and $A$ measurements -- for the MSSM spectrum.

Before considering MSSM effects in detail, it is useful to review the
general features of non-universal
CCWI contributions. Any CCWI amplitude is properly normalized to $G_\mu$,
the Fermi constant measured in
$\mu$-decay -- being one of the three most precise inputs for the gauge
sector of the MSSM. It is related
to the universal weak coupling $g$  and the $W$-boson mass as
\begin{equation}
\label{eq:Gfermi}
{G_\mu\over\sqrt{2}} = {g^2\over 8 M_W^2}\left[1+\Delta r_\mu\right]\ \ \ ,
\end{equation}
where $\Delta r_\mu$ includes the effects of weak, radiative corrections in
the MSSM as well as other possible
new physics contributions to $\mu$-decay. The MSSM amplitudes for other CCWI
processes are $\propto
g^2/M_W^2$, and in order to express them in terms of $G_\mu$, one must
invert relation (\ref{eq:Gfermi}).
For example, the Fermi constant relevant for light quark $\beta$-decay has
the form
\begin{equation}
\label{eq:Gfermibeta}
G_F^\beta = G_\mu V_{ud}(1-\Delta r_\mu+\Delta r_\beta)\ \ \ ,
\end{equation}
where $\Delta r_\beta$ contains the MSSM electroweak radiative corrections and
other possible new physics
contributions to the semileptonic decay amplitude.

The difference $\Delta r_\beta - \Delta r_\mu$ is sensitive to
non-universal effects.
Modifications of the W-boson propagator, which are universal,
cancel from this difference,
leaving only sensitivity to non-universal vertex corrections, box
diagrams, and
external leg corrections. We have computed these corrections in
the MSSM, where -- owing to
R-parity conservation -- the loops always contain an even number of
superpartners:
spin-0 sfermions (${\tilde f}$), spin-$1/2$ gluinos ($\tilde g$), and
spin-$1/2$
mixtures of electroweak gauginos and Higgsinos -- the
neutralinos (${\tilde\chi}^0_{1-4}$) and charginos
(${\tilde\chi}^+_{1,2}$). Because it compares these
corrections as they appear in leptonic and semileptonic decay amplitudes,
$G_F^\beta$ is essentially a
measure of slepton-squark universality in the MSSM. Moreover, since both
$\Delta r_\beta$ and $\Delta
r_\mu$ pertain to processes with $e^+\nu_e$ ($e^-{\bar\nu}_e$) in the final
state, the difference $\Delta r_\beta-\Delta r_\mu$ is
considerably more sensitive to effects produced by second generation
sleptons than to those produced by
the first generation. In a similar way, the ratio $R_{e/\mu}$ compares
different leptonic final states for
a given hadronic initial state, making it effectively a probe of ${\tilde
e}$-${\tilde\mu}$ universality. At present, the experimental error bars in
$R_{e/\mu}$ are roughly a factor of two larger than necessary to produce
significant constraints, so we focus on the more restrictive implications of
light quark $\beta$-decay data.

In the limit of unbroken SUSY, the parameters of the MSSM are those of the
SM, modulo a few consequences
of supersymmetry: the MSSM Lagrangian contains two Higgs doublets
($H_u$, $H_d$) and one new dimensionful parameter ($\mu$) as compared to
the SM\cite{Haber}. In order to break SUSY,
thereby splitting the masses of the SM particles from their superpartners,
one must introduce
a \lq\lq soft" SUSY-breaking Lagrangian, ${\cal L}_{\mbox{soft}}$, whose
dimensionful parameters are at
most a few $\times$ the weak scale. As a result, the MSSM contains 105 new
parameters not present in the
SM. A number of scenarios have been proposed for
simplifying ${\cal L}_{\mbox{soft}}$ into a more
fundamental theory\cite{Kane}. The task for phenomenlogy is to determine
which of these proposals is most consistent
with the data and which additional measurements could provide new
constraints on soft SUSY-breaking.

For this purpose, it is useful to identify the independent parameters which
must be determined. In the
electroweak gauge and Higgs sectors, one has the couplings $g$ and $g'$,
the vacuum expectation values of
the Higgs $v_u$ and $v_d$, the SUSY mass parameter $\mu$, along with the
SUSY-breaking  Higgs mass parameters,
$m_{H_u}^2$, $m_{H_d}^2$, $b$ and gaugino masses, $M_1$ and $M_2$. In
general, $g$,
$g'$, and
$v=\sqrt{v_u^2+v_d^2}$ can be
determined from $\alpha$, $G_\mu$, and $M_Z$, while $|\mu|$ and $b$ and be
expressed in terms of $M_Z$,
$m_{H_u}^2$, $m_{H_d}^2$ and $\tan\beta=v_u/v_d$. The strong gauge sector
of the
theory contains the SU(3)$_c$ gauge coupling, CP-violating
$\theta$-parameter, and
gluino mass,
$M_3$.
Inclusion of fermions and their superpartners introduces the SUSY Yukawa
couplings to the Higgs fields as
well as  SUSY-breaking quadratic sfermion mass terms and tri-boson
Higgs-sfermion-sfermion interactions.
As a consequence of these interactions, left- and right-handed sfermions
(${\tilde f}_{L,R}$) can mix, much as mass terms in the SM Lagrangian lead
to mixing between ordinary
fermions. The sine of the L-R mixing angle $\theta_{LR}$ is proportional to
the off
diagonal element of the
sfermion mass-squared matrix:
\begin{equation}
\label{eq:lrmix1}
M_{LR}^2 = \cases{m_f(\mu\tan\beta-A_f), &\ \ $q_f<0$;\cr
 	                 m_f(\mu\cot\beta-A_f), & \ \ \ $q_f>0$}\ \ \ ,
\end{equation}
where $m_f$ and $q_f$ are the fermion mass and charge, respectively, $A_f$
is the SUSY-breaking tri-boson
coupling, normalized to $m_f$ as in
minimal supergravity (mSUGRA) models.

To demonstrate the impact of CCWI observables on the MSSM parameter space,
it is
useful to  distinguish contributions to the corrections $\Delta
r^{\mbox{mssm}}$
arising from the SM particles ($\Delta r^{\mbox{sm}}$) and their superpartners
($\Delta r^{\mbox{susy}}$), where the former include contributions from the two
Higgs doublets. In computing $\Delta r^{\mbox{mssm}}$, we follow the standard
practice of neglecting terms quadratic in light lepton and quark masses. To
avoid
large flavor-changing neutral currents, we
also assume the same pattern of generation mixing among sfermions as for
the SM fermions.
In treating strong interaction contributions to
the semileptonic amplitudes, we adopt the framework of an effective field
theory:
short distance effects ($p>\Lambda$, $\Lambda\sim 1$ GeV) are included
explicitly in renormalized operators while long distance ($p<\Lambda$),
non-perturbative effects are contained in $\Lambda$-dependent
matrix-elements. The
latter have been studied extensively elsewhere\cite{hardy,keiser}, and we
do not
discuss them here.

We begin our analysis of the parameter space by taking
each superpartner species (${\tilde\chi}$, ${\tilde\mu}$, ${\tilde\nu_\mu}$
${\tilde
u}$, ${\tilde d}$, ${\tilde g}$) to have at least one light member whose
mass is given by the collider lower bound. In addition, we assume identical
masses and
mixing angles for
${\tilde u}$ and ${\tilde d}$ (\lq\lq squark universality"). Doing so
suppresses gluino
loop corrections to the hadronic vector current; we analyze these effects
separately
below. Under these assumptions, the nuclear $G_F^\beta$ determinations favor
maximal mixing among sfermions, with larger mass splittings between smuons
than between squarks. To
illustrate,  we plot in Fig. 1 the CCWI constraints on mass ratios
$\kappa_{\tilde f}\equiv
M_{\tilde f_2}/M_{\tilde f_1}$ for smuons and squarks. The constraints
assume maximal mixing,
which occurs for  $M_{LR}^2>> |M_L^2-M_R^2|$, where $M_{L,R}^2$ are the
diagonal mass-squared matrix entries.
The dashed line indicates the constraints obtained from $G_F^\beta$, where
we have required
that non-universal SUSY corrections to this process produce no additional
deviation from CKM unitarity.
This requirement implies that $\Delta r_\beta^{\mbox{susy}}-
\Delta r_\mu^{\mbox{susy}}<0$ (95\%
confidence).

The results indicate that the relative splitting
between charged smuons must be greater than that among squarks in order to
avoid exacerbating the CKM
unitarity deviation. This constraint may be understood by considering the
asymptotic expression for the
non-universal corrections, which is dominated by vertex corrections
involving ${\tilde
f}$-${\tilde\chi^+}$-${\tilde\chi^0}$ intermediate states. For large
$\kappa_{\tilde f}$, one has for the dominant SUSY contribution
\begin{equation}
\label{eq:asym1}
\Delta r_\beta^{\mbox{susy}}-\Delta r_\mu^{\mbox{susy}} \sim
{\alpha(c^2-s^2)\over
32\pi s^2 c^2}\ln\left(\kappa_{\tilde
q}^2/\kappa_{\tilde\mu}^4\right)+\cdots\ \ \ ,
\end{equation}
where $c$ ($s$) denotes the cosine (sine) of the weak mixing angle.
In order for this correction to be negative, the squark effect must be
smaller than the
slepton effect. In particular, for nearly degenerate squarks ($\kappa_{\tilde
q}\sim 1$), one needs $M_{\tilde\mu_2}\gtrsim 3 M_{\tilde\mu_1}$. Note that the
presence of
significant L-R mixing obscures the expected $(M/{\tilde M})^2$ scaling of the
corrections, yielding instead a logarithmic dependence on superpartner mass
ratios.

For comparison, we also give the constraints obtained from a comparsion of
the experimental value for
$G_\mu$ with its MSSM prediction using $\alpha$, $M_Z$, and $M_W$ as inputs.
In using Eq.
(\ref{eq:Gfermi}), one has
\begin{equation}
\label{eq:GfermiSM}
G_\mu^{\mbox{mssm}} = {\pi\alpha M_Z^2\over\sqrt{2}
M_W^2(M_Z^2-M_W^2)[1-\Delta r_\mu^{\mbox{mssm}}]}\ \ \ .
\end{equation}
Requiring $G_\mu^{\mbox{mssm}}$ to be consistent with the experimental value
implies the $2\sigma$ constraint
\begin{equation}
-0.0098 \leq \Delta r_\mu^{\mbox{susy}} \leq 0.0046 \ \ \ ,
\end{equation}
where the range is determined primarily by the
experimental uncertainties in $M_W$ and $m_t$.
The correction $\Delta r_\mu^{\mbox{susy}}$ samples both non-universal
corrections
as well as universal corrections entering the
$W$-boson propagator. The resulting constraints, indicated by the solid
curve in Fig.
1, lead to upper bounds on the degree of non-degeneracy among
sfermions. We observe that range of allowed mass splittings among sfermions
is tightened when the
light-quark $\beta$-decay and $W$ mass measurements are combined.

SUSY contributions to the muon anomaly are also sensitive to smuon mixing,
with the ${\tilde\mu}$-
${\tilde \chi}^0$-${\tilde\chi}^0$ vertex correction also being
proportional to $M_{LR}^2$. For light
${\tilde\mu}$ and ${\tilde\chi}^0$, the recent E821 results imply nearly
minimal mixing. Reducing the degree of smuon mixing, however, reduces the
region of the $\kappa_{\tilde\mu}$-$\kappa_{\tilde q}$ parameter space
allowed by the
CCWI data and $W$ mass measurements. For nearly minimal
mixing as implied by the muon anomaly, there exists no allowed region.
Consequently, one
must relax the assumptions that each species of superpartner has at
least one light member or that up- and down-squark masses and mixing angles are
identical. We consider each possibility in turn:

1. Increase the mass of the lightest ${\tilde\chi^0}$ and ${\tilde
\chi^+}$.  Doing so
weakens the $(g-2)_\mu$ constraints on smuon mixing. For sufficiently
heavy ${\tilde\chi^{0,+}}$, however, the effect of the dominant vertex
correction
to $G_F^\beta$ is also suppressed, leading to a change of sign in the
difference
$\Delta r_\beta^{\mbox{susy}} -\Delta r_\mu^{\mbox{susy}}$ (the
contribution from other
vertex corrections and box graphs is positive). In order to keep this
difference
negative, one must increase $\kappa_{\tilde\mu}$, leading to a correlation
between
$\kappa_{\tilde\mu}$ and $M_{\tilde\chi^0}$. For
$M_{\tilde\chi^0} < 1$ TeV, the allowed regions determined by the
$(g-2)_\mu$ and $G_F^\beta$ constraints never overlap.

2. Increase the mass of the lightest ${\tilde\mu}$ (and, thus,
${\tilde\nu_\mu}$) for
fixed $M_{{\tilde u},{\tilde d}}$ and $\kappa_{\tilde\mu}$. While this
parameter
variation desensitizes the muon anomaly to smuon mixing, it also reduces
the region
allowed by the comparison of $G_\mu^{\mbox{mssm}}$ with the experimental
value. When
$M_{\tilde\mu}$ is sufficiently heavy to evade the
$(g-2)_\mu$ constraints on smuon mixing, the allowed region of Fig. 1 vanishes.
Increasing the mass of the lightest first generation squarks does not affect
$(g-2)_\mu$, leaving  the essential conflict between the muon anomaly and
CCWI data (Fig. 1) unresolved.

3. Take smuon mixing to be zero, thereby avoiding any
$(g-2)_\mu$ constraints on $M_{LR}^2$. Since the degree of smuon mixing must
always be greater than mixing among first generation squarks (Fig. 1), we
also take
the latter to zero. In this case, the
$G_F^\beta$ constraints lead to a non-trival relationship between the masses of
the ${\tilde\mu}_L$ and first generation ${\tilde q}_L$. In particular,
one always has $M_{\tilde\mu_L}>M_{\tilde q_L}$ for light
${\tilde\chi}$ and no sfermion mixing. In this case, the dominant MSSM
correction
to $G_F^\beta$ has the asymptotic (large $M_{\tilde f}$) expression
\begin{eqnarray}
\label{eq:asym2}
\Delta r_\beta^{\mbox{susy}}-\Delta r_\mu^{\mbox{susy}}&\sim& {\alpha\over
2\pi}\cos
2\beta\times\\
\nonumber
[{1\over 3}{M_Z^2\over M_{\tilde q}^2}
\ln{M_{\tilde q}^2\over <M_{\tilde\chi}^2>}
&-&{M_Z^2\over M_{\tilde\mu}^2}
\ln{M_{\tilde\mu}^2\over <M_{\tilde\chi}^2>}]+\cdots
\ \ \ .
\end{eqnarray}
where $<M_{\tilde\chi}^2>$ is the mass scale associated with the ${\tilde\chi}$
which is much smaller than $M_{\tilde f}$ in the asymptotic limit.
For $\tan\beta>1$ as favored by lower bounds on the lightest Higgs
mass\cite{Hei00},
$\cos 2\beta<0$. To maintain the correct sign for the SUSY contribution to
$G_F^\beta$,
one requires $M_{\tilde\mu}^2 > 3M_{\tilde q}^2$ (up to logarithmic
corrections),
where the factor of three results from the different hypercharges of the
doublet
left-handed squarks and smuons.

This phenomenological solution is particularly interesting from
the standpoint of both gauge-mediated and mSUGRA models of SUSY-breaking
mediation,
which  generally predict $M_{\tilde q} > M_{\tilde\ell}$. In mSUGRA, this
hierarchy
results from  gluino contributions to the renormalization group running of the
masses down from the GUT
scale. Inverting this hierarchy would presumably require modifying the
universality
assumptions made for the parameters of ${\cal L}_{\mbox{soft}}$ at the GUT
scale.

4. Relax the assumption of squark universality. For illustrative purposes, we
take no
mixing among ${\tilde u}_{L,R}$ but maximal mixing among
${\tilde d}_{L,R}$ and ${\tilde\mu}_{L,R}$ (similar qualitative conclusions
arise if
mixing occurs among
${\tilde u}_{L,R}$ instead of ${\tilde d}_{L,R}$). In this case, SUSY SU(3)$_c$
corrections to the light quark current operator dominate $\Delta
r_\beta^{\mbox{susy}}$
for light
gluinos. The resulting
$\kappa_{\tilde\mu}$-$\kappa_{\tilde d}$ constraints are shown in Fig. 2.
The solution
in this instance requires significant mixing and mass-splittings among the
d-type
squarks. For $\kappa_{\tilde\mu}\sim 1$ as required by $(g-2)_\mu$, we require
$3\lesssim \kappa_{\tilde d} \lesssim 4.5$. Decreasing the degree of ${\tilde
d}_{L,R}$ mixing reduces this range. For $M_{\tilde d_1}\sim 115$ GeV,
the lower bound of this range implies $M_{\tilde d_2}\sim 450$ GeV, or
$M_{LR}^2/m_d=
(\mu\tan\beta-A_d)\sim 10^7$ GeV. Increasing the mass of the lightest
${\tilde d}$
(see, {\em e.g.}, \cite{erler}) leads to a corresponding increase in this
scale. Short
of any miraculous fine tuning of parameters involving smuons, the
$(g-2)_\mu$ results cannot accomodate values of $\mu
\tan\beta$ having this magnitude, so the entire effect would have to arise
from $A_d$.
In models such as mSUGRA, the tri-boson couplings $A_f$ have a common value at
the GUT scale. Lower bounds on the mass of the lightest Higgs imply that
$|A_t| \lesssim $
a few TeV at the weak scale\cite{Hei99}, while perturbative running of the
$A_f$ down
from a common value at the GUT scale cannot produce a factor of $10^4$
difference
between $A_d$ and $A_t$ at the weak scale. Thus, it appears that this solution
cannot be accomodated within SUSY-breaking mediation models
assuming coupling universality at the GUT scale.

The implications of solution 4 may be evaded for sufficiently heavy gluinos
($M_{\tilde g}\gtrsim 500$ GeV). In
this case, however, one returns to the situation characterized by Fig. 1,
leaving
solution 3 -- which is also inconsistent with the universality of soft
SUSY-breaking
interactions -- as the only viable option.

Alternatively, one may avoid either of these
solutions by taking {\em all} superpartners to be  sufficiently heavy
(${\tilde M}\gtrsim 500$ GeV),
a choice disfavored by CCWI and $(g-2)_\mu$ data.
Indeed, while MSSM loop effects can never completely correct the apparent CKM
unitarity violation suggested by $G_F^\beta$, choosing MSSM parameters near
the edges
of the allowed regions (for light superpartners) can reduce the discrepancy
by $\sim
0.5\sigma$. More generally,
requiring consistency between the MSSM and both CCWI and $(g-2)_\mu$ data
suggests the
necessity of modifying widely-used models of SUSY-breaking mediation.

We thank J. Erler, B. Filippone, S. Heinemeyer, John Ng, and P. Vogel for
several useful
discussions. This work was supported by the Department of Energy and
National Science
Foundation.




\begin{figure}
\hspace{0.1in}
\includegraphics[height=2.25in,width=2.5in]{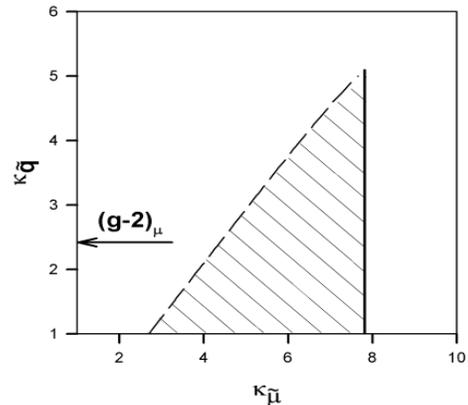}
\caption{\label{Fig1}Slepton-squark universality constraints for aligned first
generation squarks with maximal mixing. Shading indicates region allowed by
$G_F^\beta$ (dashed line) and $G_\mu$-$\alpha$-$M_Z$-$M_W$ relation (solid
line). Arrow indicates $(g-2)_\mu$ allowed region ($1<\kappa_{\tilde\mu}\leq 1.04$).} \end{figure}  \begin{figure} \hspace{0.05in}
\includegraphics[height=2.25in]{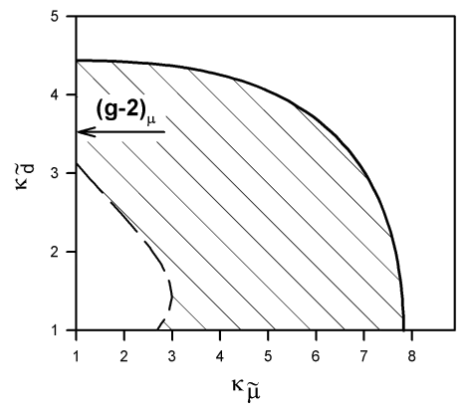}
\caption{\label{Fig2}Same as Fig. 1 but for squark non-universality.}
\end{figure}


\begin{thebibliography}{99}

\bibitem{barger}V. Barger and K. Cheung, Phys. Lett. B480, 149 (2000).

\bibitem{MRM00}M.J. Ramsey-Musolf, Phys. Rev. D62:056009 (2000).

\bibitem{dreiner}B.C. Allanach, A. Dedes, and H. Dreiner, Phys. Rev. D60:075014
(1999).

\bibitem{Brown} Muon g-2 Collaboration, H.N. Brown {\em et al.}, Phys. Rev.
Lett.
{\bf 86}, 2227 (2001).

\bibitem{Everett} For a recent discussion and reference list, see, L. Everett
{\em et al.}, Phys. Rev. Lett. {\bf 86}, 3484 (2001).

\bibitem{hardy}I.S. Towner and J.C. Hardy, nucl-th/9809087.

\bibitem{Haber}H.E. Haber and G.L. Kane, Phys. Rep. 117, 75 (1985).


\bibitem{Kane} For a review, see Perspectives on Supersymmetry, G.L. Kane, Ed.,
World Scientific, Singapore, 1998.

\bibitem{keiser} N. Kaiser, Phys. Rev. C 64:028201 (2001).


\bibitem{Hei00} S. Heinemeyer and G. Weiglein, Nucl. Phys. B (Proc. Suppl.)
89, 216
(2000).

\bibitem{erler} J. Erler and D. M. Pierce, Nucl. Phys. B 526, 53 (1998).

\bibitem{Hei99} H. Heinemeyer, W. Hollik, and G. Weiglein, Phys. Lett. B
455, 179
(1999).

\end{thebibliography}
\end{document}